\documentclass[aps,showpacs,manuscript,12pt]{revtex4}
\usepackage{amssymb}
\usepackage{amsmath}
\usepackage{graphicx}

\begin{document}

\title{Non-linear gyrokinetic theory of magnetoplasmas}
\author{Massimo Tessarotto$^{1,2}$, Claudio Cremaschini$^{1}$, Piero Nicolini$^{1}$ and Alexei Beklemishev$^{3}$}
\affiliation{$^{1}$Department of Mathematics and Informatics,University of Trieste, Trieste, Italy \\
$^{2}$ Consortium for Magnetofluid Dynamics, Trieste, Italy\\
$^{3}$ Budker Institute of Nuclear Physics, Novosibirsk, Russia}

\begin{abstract}
A crucial issue in relativistic plasma, particularly relevant in
the astrophysical context, is the description of highly magnetized
plasmas based on a covariant formulation of gyrokinetic dynamics.
An interesting case in question is that in which the background
electric field (produced either by the same plasma of by other
sources) results suitably small (or vanishing) with respect to the
magnetic field, while at the same time short-wavelength EM
perturbations can be present. The purpose of this work is to
extend the
relativistic gyrokinetic theory developed by Beklemishev \textit{et al. }%
[1999-2005] to include, in particular, also the treatment of such
a case. We intend to show that this requires the development of a
perturbative expansion involving simultaneously both the particle
4-position vector and the corresponding \textit{4-velocity
vector}. For this purpose a synchronous form of the relativistic
Hamilton variational principle is adopted.
\end{abstract}

\maketitle


\section{1. Introduction}

This work is a part of a systematic investigation concerning the
relativistic gyrokinetic theory developed by Beklemishev \textit{et al.}\cite%
{Beklemishev1999,Beklemishev2004,Nicolini2005} and its
applications. \ In this paper, in particular, we intend to propose
an extended formulation of covariant gyrokinetic theory
specifically intended to include the effect of 4-velocity
perturbations, as appropriate for the description of relativistic
magnetoplasmas in the presence of suitably "small" ( and regular,
but otherwise arbitrary) electromagnetic and gravitational
perturbations produced, via collective interactions, by the same
plasma. The physical motivations of the investigation are related
to the occurrence of strongly magnetized relativistic plasmas in
curved space-time. In the astrophysical context these plasmas are
typically related to the existence of accretion disks, plasma
inflows and outflows and relativistic jets, all occurring
typically close to massive object, such as neutron stars, black
holes (BH) and active galactic nuclei (Mohanty \cite{Mohanti}).
These plasmas can be characterized by the presence of ultra-strong
magnetic fields, possibly generated by the same plasma. Although,
direct measurements of $\mathbf{B}$ can be very difficult, for
example in jets there is yet no reliable observation of the
magnetic field intensity, the partial information already
available for other cases indicates that in suitable circumstances
the magnetic field can become extremely intense. For example,
sufficiently close to the surface of pulsars the magnetic field
intensity can reach values well
above the so-called quantum critical field strength of $B_{c}={m_{e}^{2}}{%
c^{3}/e\hbar=4.4x10^{13}G}$. Magnetic fields of this order or
larger are said to be super strong. A similar situation might
occur for BH's surrounded by strong magnetic fields, and in
particular in the regions of space-time suitably close to the
so-called photon-sphere \cite{Nicolini2006}, where collimated
outflows can arise.

While such fields are typically expected to manifest together with
intense electric fields ($\mathbf{E}$), there can be cases in
which there results in particular
\begin{equation}
E/B\ll 1  \label{1}
\end{equation}%
(\textit{weak electric field assumption}). Here $E$ and $B$ are
the invariant eigenvalues of the electromagnetic (EM) field tensor
$F_{\mu \nu }$ (see below). This may happen, for example, in
plasmas in which the magnetic field is produced primarily by
strong diamagnetic currents. These, by definition, are electric
currents flowing in the plasma which are driven by gradients of
the relevant fluid fields (in particular, fluid velocity, number
density, temperature, magnetic field intensity, etc.) and,
instead, are independent of the (local) electric field.

An example is provided by plasma outflows (or jets) in which the
kinetic energy of the escaping particles is usually believed to by
provided by the
presence of a quasi-stationary magnetic field moving with the plasma \cite%
{Blanford}. In all such phenomena EM turbulence is expected to be
ubiquitous and to affect in a significant way plasma dynamics. The
region where these processes may act is typically too close to the
central object (the BH or a neutron star) to be resolved
observationally. Thus, one possible way to obtain some insight
into this region is by studying the properties of these phenomena.
In particular, the emitted radiation spectrum could be
significantly modified by the effect of a finite curvature of
space-time, as well as by short-wavelength EM and gravitational
perturbations in the region of emission\cite{Frank1995}.

An important theoretical problem is therefore the prediction of
such effects based on single-particle gyrokinetic dynamics. Such
phenomena require their description in the framework of a
consistent covariant theory taking into account the effect of
space-time non-uniformity as well as the possible presence of
wave-fields, in principle, both EM and gravitational. A convenient
formulation is provided by the covariant gyrokinetic theory
recently developed by Beklemishev \textit{et al.} \cite%
{Beklemishev1999,Beklemishev2004}. As well known, object of
particle gyrokinetics is the description of particle dynamics in
terms of a reduced set of variables (so-called gyrokinetic). These
include, in particular, the
fast gyration angle around the magnetic flux lines, i.e., the gyrophase $%
\phi $, which results ignorable. Gyrokinetic theory is therefore
realized by the construction of a suitable, generally
non-canonical, phase-space transformation. From a mathematical
viewpoint, gyrokinetic theory provides a perturbative description
of particle dynamics in terms of \textquotedblleft
simpler\textquotedblright\ equations of motion, which are
characterized by a reduced number of effective degrees of freedom.
Goal of the present paper is to extend the Beklemishev and
Tessarotto formulation to obtain the gyrokinetic theory in curved
background space-time which applies also to the special case
\ref{1} and includes the simultaneous effect of short-wavelength
EM and gravitational perturbations. Starting point is the
Larmor-radius ordering, requiring $\varepsilon \sim
\frac{r_{L}}{L_{EM}}\sim
\frac{r_{L}}{L_{g}}\ll 1,$and $\lambda \equiv \frac{\lambda _{EM}}{L_{EM}}%
,\delta \equiv \frac{\lambda _{g}}{L_{g}}\ll 1,$where $r_{L}$ is
the Larmor radius and $L_{EM},L_{g},$respectively denote the
characteristic scale lengths of the background and wave EM and
gravitational fields. In
particular, we intend to determine new expressions for the magnetic moment $%
p_{\phi }\equiv \widehat{\mu }$ and of the cyclotron frequency $\frac{d}{ds}%
\phi $, possibly modified by the effect of space-time curvature.

\section{2. The synchronous hybrid Hamilton variational principle in curved
space}

The starting point for the construction of the gyrokinetic theory
in curved space-time is the proper treatment of the variational
principle which determines the gyrokinetic world lines $\left\{
r^{\prime \alpha }(s),s\in I\right\} $ in the presence of a
non-uniform metric tensor $g_{\mu \nu
}(r^{\alpha }),$as well as the EM 4-vector potential $A_{\mu }$. Here $%
r^{\prime \alpha }$ denotes the gyrocenter position 4-vector (in
the sequel the apex will denote gyrokinetic variables or
phase-functions evaluated at
the gyrocenter position). \ Historically (see for example Einstein \cite%
{Einstein}, Landau \& Lifschitz \cite{LL1}) the relativistic
equations of motion were obtained using an asynchronous
variational principle, stemming from the minimum-action principle
(relativistic Hamilton principle). Accordingly, the trajectory of
a point-particle is described as an extremal curve (i.e., a
geodesics) in the 4-dimensional space-time. The choice of an
asynchronous formulation, rather than an asynchronous one, is a
natural consequence of the fact that the variational principle is
a generalization of the well-known Hertz-Fermat-Maupertius
variational principle (Goldstein
\textit{et al.} \cite{Goldstein}) and that the 4-velocity of a particle, $%
u^{\mu }(s)=\frac{d}{ds}r^{\mu }(s)$ , is a unit 4-vector, i.e.,
it must
fulfill the constraint $u_{\mu }u^{\mu }=1$ (\textit{realizability condition}%
)$,$ where $u_{\mu }=g_{\mu \nu }u^{\nu }.$ As a consequence, it
follows that the line element $ds$ along the world line $\left\{
r^{\mu }(s),s\in I\right\} $ must satisfy the constraint
$ds^{2}=g_{\mu \nu }dr^{\mu }(s)dr^{\nu }(s),$which implies that
the asynchronous variation of $ds$ results $\Delta \left(
ds\right) \neq 0,\Delta $ denoting the operator of asynchronous
variation. \ On the other hand, adopting the [relativistic]
asynchronous principle one has to face the conceptual difficulty
represented by the fact that - contrary to what happens for the
non-relativistic Hamilton principle - it cannot provide a
variational principle for the relativistic Hamilton equations (at
least using a Legendre transformation like in non-relativistic
mechanics). In reality, the difficulty can be circumvented (see
for example Wheeler \textit{et al.}\cite{Wheeler} and Goldstein
\textit{et al. }\cite{Goldstein}) adopting a suitable synchronous
variational principle. In such a case it is immediate to show that
the canonic form of the relativistic Hamilton equations is
recovered, just like in non-relativistic mechanics, by means of a
suitable Legendre transformation. Finally, it must be recalled
that variational principles, both synchronous and asynchronous,
can be expressed in arbitrary "hybrid" (i.e., generally
non-canonical) variables, provided these variables are connected
by a diffeomorphism. to the original one [i.e., for example to be
identified with the canonical variables], thus yielding a
\textit{hybrid variational principle}. These variable can also be
superabundant, since the hybrid variables can be defined in such a
way to span, in principle, phase-spaces of arbitrary dimension.
Thus, they can be described in the
framework of the so-called \textit{constrained dynamics} (Dirac \cite{Dirac}%
), in turn relying on the well-known \textit{method of Lagrange multipliers}%
. In particular, as a special choice, the hybrid variables can
also be identified with appropriate gyrokinetic variables. In this
section we intend to obtain a synchronous\ variational principle
of this type, in which the hybrid variables are provided by a
suitable set of superabundant
non-canonical variables.\ For definiteness let us introduce the action $S(%
\mathbf{y})\equiv \int_{_{1}}^{2}\gamma (\mathbf{y})+dF,$where
$\mathbf{y}$ denotes the set of variables $\mathbf{y\equiv }\left(
r^{\mu },u^{\mu
},\lambda \right) ,$ $\gamma $ a suitable differential 1-form and $F(\mathbf{%
y},s)$ an arbitrary gauge function [i.e., a real function which is at least $%
C^{(2)}$ in its existence domain]. In particular, let us impose
that the fundamental differential 1-form $\gamma $ takes the value
\begin{equation}
\gamma (\mathbf{y})=g_{\mu \nu }\left[ qA^{\nu }+u^{\nu }\right]
dr^{\mu }+\lambda \left[ g_{\mu \nu }u^{\mu }u^{\nu }-1\right]
\label{diff 1-form  A}
\end{equation}%
Here the notation is standard \cite{Beklemishev1999,Beklemishev2004}. Thus, $%
A^{\nu }$ are the counter-variant components of the EM 4-potential and $%
q=Ze/\ m_{o}c$ is the normalized charge. and, moreover, $r^{\nu }$ and $%
u^{\nu }$ coincide with the countervariant components of the 4-vectors $%
\mathbf{r}$ (4-position) and $\mathbf{u}$ (4-velocity). It is
immediate to
show that phase-space trajectory of the point particle with 4-position $%
r^{\nu }$ and 4-velocity $u^{\nu }$ results, by construction, an
extremal curve of the variational functional $S(\mathbf{y})$.
Another equivalent possibility is to represent the differential
1-form in terms of
\begin{equation}
\gamma (r^{\mu },u_{\mu },\lambda )=\left[ qA_{\mu }+u_{\mu
}\right] dr^{\mu }-\frac{1}{2}\left[ g^{\mu \nu }u_{\mu }u_{\nu
}-1\right] , \label{diff 1-form  B}
\end{equation}%
where the Lagrange multiplier $\lambda $ has been set equal to its
extremal
value, $\lambda =-1/2$. Finally, let us require, in case (\ref{diff 1-form A}%
), that $S$ be defined in the synchronous functional class of real
functions
\begin{eqnarray}
&&\left. \left\{ \mathbf{y}\right\} \equiv \left\{
\mathbf{x}(s)\equiv \left( r^{\mu }(s),u^{\mu }(s),\lambda
(x)\right) :\mathbf{x}(s)\in
C^{(2)}(I),\right. \right. \\
&&\left. \mathbf{x}(s)\in \mathbb{R}^{9},I\equiv \left[
s_{1},s_{2}\right] \subseteq
\mathbb{R},\mathbf{x}(s_{i})=\mathbf{x}_{1}(i=1,2)\right\} ,
\notag
\end{eqnarray}%
while for (\ref{diff 1-form B}), $\lambda $ is considered prescribed ($%
\lambda =-1/2$) It is immediate to prove that \textit{in both cases} (\ref%
{diff 1-form A}) and (\ref{diff 1-form B}) the variational
principle
\begin{equation}
\delta S=0  \label{variational principle}
\end{equation}%
delivers the correct relativistic equations. We remark that the
synchronous variations $\delta r^{\mu },\delta u^{\mu },\delta
\lambda $ must be all considered independent. Indeed, the physical
realizability condition is satisfied \textit{only} by the extremal
curve. Hence, in a proper sense, this represents an example of
unconstrained variational principle.

\section{3. Extended gyrodynamics}

To obtain the new extended formulation of gyrokinetic theory, the
variational functional $S(\mathbf{y})$ must be expressed in terms
of gyrokinetic variables, which requires a suitable
\textit{extended} phase space transformation to a new set of
variables\ $y^{i},$\thinspace including in particular the
gyrophase $\phi $ which describes the fast gyration motion of
particles around magnetic flux lines. Usually such a
transformation is obtained by means of perturbation theory in
$\varepsilon ,$ to be considered infinitesimal and defined as the
ratio between the Larmor radius and the smallest characteristic
scale length of the magnetoplasma$.$ In particular
this requires that the EM $4-$potential $A^{\mu }$ and the metric tensor $%
g_{\mu v}$ are assumed to be expressed in terms of the
representations
\begin{eqnarray}
A^{\mu } &=&\frac{1}{\varepsilon }A^{\mu }(r^{\alpha })+\xi a^{\mu
}(r^{\alpha }/\xi ),  \label{pert-1} \\
g_{\mu v} &=&G_{\mu \nu }(r^{\alpha },\varepsilon )+\xi g_{\mu
v}^{1}(r^{\alpha }/\xi )  \label{pert-2}
\end{eqnarray}%
(\textit{Larmor-radius ordering}). Here $\frac{1}{\varepsilon
}A^{\mu }(r^{\alpha },\varepsilon )$ and $G_{\mu \nu }(r^{\alpha
},\varepsilon ),$ denoting the equilibrium terms, are assumed to
be $C^{(\infty )}$ in all variables, while $\xi a^{\mu }(r^{\nu
}/\xi )$ and $\xi g_{\mu v}^{1}(r^{\nu }/\xi ),$ define the
short-wavelength (and high-frequency) perturbations. In
particular, $\xi $ is an infinitesimal parameter, assumed
generally independent of $\varepsilon ,$ and in particular such
that $\varepsilon \ll \lambda ,$ which characterizes the tensors
$\xi a^{\mu }(r^{\nu }/\xi )$ and $\xi g_{\mu v}^{1}(r^{\nu }/\xi
).$Hence, the equilibrium quantities can be Taylor expanded near
$r^{\prime \alpha }$
\begin{align}
\frac{A^{\mu }}{\varepsilon }& =\frac{A^{\prime \mu }}{\varepsilon }%
+A_{1}^{\mu }+o(\varepsilon ), \\
G_{\mu v}& =G_{\mu \nu }^{\prime }+\varepsilon G_{\mu
v}^{1}+o(\varepsilon ),
\end{align}%
where $A^{\prime \mu }=A^{\mu }\left( r^{\prime \alpha }\right) ,$
$G_{\mu \nu }^{\prime }=g_{\mu \nu }\left( r^{\prime \alpha
}\right) ,$ $A_{1}^{\mu }=r_{1}^{\alpha }\partial _{\alpha
}^{\prime }A^{\prime \mu }$ and $G_{\mu v}^{1}\equiv r_{1}^{\alpha
}\partial _{\alpha }^{\prime }G_{\mu v}^{\prime }. $ The
perturbations of relevant fields depend generally on the gyrophase
angle $\phi ^{\prime }$ through $\varepsilon r_{i}^{\alpha }\left(
y^{i}\right) .$ The perturbative scheme converges, at least in an
asymptotic sense, only if one assumes the validity of a suitable
ordering conditions (Larmor radius ordering), implying that EM
contributions must dominate, in a suitable sense, over inertial
ones in the variational functional. Formally this corresponds to
replace the charge $e$ by $e/\varepsilon ,$ leaving the rest mass
$m_{0}$ unchanged. In this work, extending the approach developed
by Beklemishev \textit{et al.}\cite%
{Beklemishev1999,Beklemishev2004,Nicolini2005}, the gyrokinetic
transformation is taken of the general form $\left( r^{\alpha
},u^{\beta }\right) \leftrightarrow y^{i}\equiv \left( r^{\prime
\alpha },u^{\prime \alpha }\right) ,$ where $r^{\prime \alpha }$
and $u^{\prime \alpha }$ identify respectively the so-called
gyrocenter\ 4-vector and a transformed
4-velocity, to be determined respectively in terms of the two power series%
\begin{eqnarray}
r^{\mu } &=&r^{\prime \mu }+\sum\limits_{s=1}^{\infty }\varepsilon
^{s}r_{s}^{\mu }  \label{expansion position} \\
u^{\mu } &=&u^{\prime \mu }\oplus \sum\limits_{s=1}^{\infty }\xi
^{s}v_{s}^{\mu }(\mathbf{y})  \label{expansion velocity}
\end{eqnarray}%
(\textit{extended gyrokinetic transformation}). Here $r_{i}^{\mu
}\left( y^{i}\right) $ and $v_{i}^{\alpha }\left( y^{i}\right) $
($i=0,1,$..) are respectively the 4-position and 4-velocity
perturbations, while the operator $\oplus $ must be suitably
defined as corresponds to the 4-velocity addition law
\cite{Tessarotto1998}. In particular, the perturbations
$r_{i}^{\mu }\left( y^{i}\right) $ and $v_{i}^{\alpha }\left(
y^{i}\right) $ are defined
in such a way that both are pure oscillatory with respect to the gyrophase $%
\phi ^{\prime }.$ As usual, $u^{\prime \alpha }$ can be
characterized in terms of suitable independent variables, one of
which includes the gyrophase $\phi ^{\prime }.$ This is obtained
by projecting $u^{\prime \alpha }$ along the four independent
directions defining the so-called fundamental tetrad
unit 4-vectors $e_{a}^{\mu }$ (with $a=0,...,3),$ hereon also denoted $%
\left( l^{\prime \mu },l^{\prime \prime \mu },l^{\mu },\tau ^{\mu
}\right) .$ In particular, the latter can always be identified
with the EM fundamental tetrad, i.e., the basis vectors of the EM
field tensor admitting , in this case to be evaluated at the
gyrocenter position, $F_{\mu \nu }^{\prime }\equiv F_{\mu \nu
}(x^{\prime \alpha }),$ with eigenvalues $B$ and $E.$ In
particular, the space-like vectors $e_{2}^{\mu },e_{a3}^{\mu }$
are assumed to satisfy the eigenvalue equations
\begin{align}
F_{\mu \nu }^{\prime }e_{2}^{\nu }& =Be_{3\mu },
\label{eigenvalue equations} \\
F_{\mu \nu }^{\prime }e_{3}^{\nu }& =-Be_{2\mu }.  \notag
\end{align}%
Here $e_{a\mu }=g_{\mu \nu }^{\prime }e_{a}^{\nu }$ are the
covariant components of $e_{a}^{\mu }$ and the labels $a$ assume a
tensor meaning in the tangent space. Thus we shall raise and lower
them in terms of $e_{\mu }^{a}=\eta ^{ab}e_{b\mu },$ by means of
the Minkowskian tensor $\eta =diag\left( 1,-1,-1,-1\right) .$ It
follows to leading order in $\varepsilon $
\begin{equation}
u^{\prime \mu }=a^{\mu }\cos \phi ^{\prime }+b^{\mu }\sin \phi ^{\prime }+%
\overline{u}^{\mu }
\end{equation}%
which can be taken as a definition for the gyrophase $\phi $. Here $%
\overline{g}\equiv \left\langle g\right\rangle _{\phi }=\frac{1}{2\pi }%
\int_{0}^{2\pi }g\left( y\right) d\phi ^{\prime }$ denotes the
$\phi
^{\prime }-$average of a function $g\left( y\right) $ and $a^{\mu }$ and $%
b^{\mu }$ can be identified, up to a constant factor, with the two
space-like 4-vectors, $e_{2}^{\mu }$ and $e_{3}^{\mu }$ (which
together with $\overline{u}^{\mu }$ are evaluated at the
gyrocenter position $x^{\prime \alpha }),$ thus letting $a^{\mu
}=w$ $e_{2}^{\mu }$ and $b^{\mu
}-we_{3}^{\mu }.$ The components $a^{\mu },$ $b^{\mu }$ and $\overline{u}%
^{\mu }$ are not completely arbitrary, but satisfy the constraint
$g_{\mu \nu }u^{\mu }u^{\nu }=1$ for all $\phi ^{\prime }.$ Thus,
we shall demand the mutual orthogonality
\begin{equation}
a^{\mu }b_{\mu }=a^{\mu }\overline{u}_{\mu }=b^{\mu
}\overline{u}_{\mu }=0, \label{orthogonality}
\end{equation}%
while their moduli are expressed via a single real scalar $w$
\begin{equation}
a^{\mu }a_{\mu }=b^{\mu }b_{\mu }=1-g_{\mu \nu }\overline{u}^{\mu }\overline{%
u}^{\nu }=-w^{2}.  \label{modulus}
\end{equation}%
We stress that only to the leading order $\overline{u}_{\mu
}=g_{\mu \nu }^{\prime }\overline{u}^{\mu }$ where $g_{\mu \nu
}^{\prime }\equiv g_{\mu \nu }(x^{\prime \alpha })$ (and similarly
$a_{\mu }=g_{\mu \nu }^{\prime
}a^{\nu },$ $b_{\mu }=g_{\mu \nu }^{\prime }b^{\nu }$)$,$ while in general $%
\overline{u}_{\mu }=g_{\mu \nu }^{\prime }\overline{u}^{\nu }\neq
g_{\mu \nu
}\overline{u}^{\nu }.$ From orthogonality relations (\ref%
{orthogonality}) one finds the countervariant representation:
\begin{equation}
\overline{u}^{\mu }=u^{0}e_{0}^{\mu }+u^{\parallel }e_{1}^{\mu }.
\end{equation}%
Here $u^{0}=u_{0},$ and $u^{\parallel }$ is the "parallel" component $%
u^{\parallel }=-u_{\parallel },$ with $u_{0}$ and $u_{\parallel }$
denoting
the covariant components of \ $\overline{u}^{\mu },$ and $\mathbf{e}_{0},%
\mathbf{e}_{1}$ are the remaining time-like and space-like unit
vectors of the tetrad. From Eq. (\ref{modulus}) one obtains
$u_{0}^{2}=1+u_{\parallel }^{2}+w^{2},$ which permits to express
$u_{0}$ by means the independent functions $u_{\parallel },w$.
With the above positions for the transformed 4-velocity it can be
shown that $r_{i}^{\mu }\left( y^{i}\right) $ are
purely oscillatory arbitrary 4-vectors, namely the $\phi ^{\prime }-$%
averages of $r_{i}^{\mu }\left( y^{i}\right) $ are zero.

\section{4. Gyrokinetic equations in curved space}

The extended \ gyrokinetic transformation (\ref{expansion position}),(\ref%
{expansion velocity}) can be carried out in principle at any order in $%
\varepsilon $ and $\xi .$ Here, to illustrate some interesting
features of the approach we intend to formulate the leading-order
perturbative theory with respect to the parameter $\varepsilon $
(i.e., of order $\varepsilon ^{1}$). In this case the variational
functional $S(\mathbf{y})$ when expressed in terms of the
gyrokinetic variables $\left( \phi ^{\prime },u_{0},u_{\parallel
},\overset{\wedge }{\mu },r^{\prime \mu },\lambda \right) $ can be
shown to take the form

\begin{equation}
S_{g}(\mathbf{y}^{\prime })\equiv \int \gamma _{g}\left( \phi
^{\prime },u_{0},u_{\parallel },\overset{\wedge }{\mu },r^{\prime
\mu },\lambda \right) ,  \label{gyrokinetic functional}
\end{equation}%
where
\begin{eqnarray}
&&\left. \gamma _{g}\left( \phi ^{\prime },u_{0},u_{\parallel },\overset{%
\wedge }{\mu },r^{\prime \mu },\lambda \right) =\left\{ \left( \frac{q}{%
\varepsilon }A_{\mu }^{\prime }+u_{\parallel }l_{\mu }+u_{0}\tau
_{\mu
}\right) dr^{\mu }+\right. \right.   \label{gyrokinetic diff 1-form} \\
&&\left. +\overset{\wedge }{\mu }d\phi +\lambda \left[ u_{0}^{2}-u_{%
\parallel }^{2}-2qB\overset{\wedge }{\mu }-1\right] ds\right\}   \notag
\end{eqnarray}%
is the fundamental gyrokinetic differential 1-form. In particular, in case (%
\ref{diff 1-form B}) there results $\lambda =-1/2$ in
Eq.(\ref{gyrokinetic
diff 1-form}). Here the notation is standard and in particular $\overset{%
\wedge }{\mu }$ is the relativistic magnetic moment \cite%
{Beklemishev1999,Beklemishev2004,Nicolini2005}. Next, we wish to
investigate the consequences of the new synchronous variational
principle introduced in Sec.2, in which the . We have proven that
the Lagrange multiplier $\lambda $
must necessarily take the value $\lambda =-\frac{1}{2}$ in both cases (\ref%
{diff 1-form A}) and (\ref{diff 1-form B}), thanks to the
variational principle \ (\ref{variational principle}). The same
result manifestly holds also for the gyrokinetic functional
$S_{g}(\mathbf{y}^{\prime }),$ hence
\begin{equation}
\lambda =-\frac{1}{2}\frac{\tau _{\mu }}{u_{0}}\left( wl^{\prime
\mu }\cos (\phi )+wl^{\prime \prime \mu }\sin (\phi )+u_{\parallel
}l^{\mu }+u_{0}\tau ^{\mu }\right) =-\frac{1}{2}.  \label{135}
\end{equation}%
Therefore, the Euler-Lagrange equations for the variables $\left(
\phi ^{\prime },u_{0},u_{\parallel },\overset{\wedge }{\mu
},r^{\prime \mu
}\right) $ achieve the simple expressions%
\begin{eqnarray}
&&\left. d\overset{\wedge }{\mu }=0\Rightarrow \overset{\wedge }{\mu }%
=const.\right.   \label{136} \\
&&\left. \tau _{\mu }dr^{\prime \mu }-u_{0}ds=0\Rightarrow u_{0}=\frac{%
dr^{\prime \mu }}{ds}\tau _{\mu },\right.   \label{2} \\
&&\left. l_{\mu }dr^{\prime \mu }+u_{\parallel }ds=0\Rightarrow
u_{\parallel
}=-\frac{dr^{\prime \mu }}{ds}l_{\mu },\right.   \label{3} \\
&&\left. d\phi ^{\prime }+qBds=0\Rightarrow \frac{d\phi ^{\prime }}{ds}%
=-\Omega ,\right.   \label{4} \\
&&\frac{q}{\varepsilon }F_{\mu \nu }^{\prime }dr^{\prime \nu
}-l_{\mu }du_{\parallel }-\tau _{\mu }du_{0}-2\chi
q\overset{\wedge }{\mu }H_{,\mu
}=0,  \label{5} \\
&&\left. u_{0}^{2}-u_{\parallel }^{2}-2qB\overset{\wedge }{\mu
}-1=0,\right. \label{7}
\end{eqnarray}%
where $\Omega =qB$ is the particle gyrofrequency (or cyclotron
frequency) defined with respect to the rest mass. It is
interesting to point out that Eq.(\ref{4}) is formally identical
to the non-relativistic one. Hence, at this order \textit{no
correction due to space-time curvature} appears in this equation.
\ The remaining equations have a straightforward meaning. Thus,
Eq.(\ref{136}) implies the conservation of the relativistic
magnetic moment $\overset{\wedge }{\mu },$which is therefore an
adiabatic invariant.
Actually the result is expected to hold at any order in $\varepsilon $ and $%
\lambda ,$ although the perturbative theory locally may not
converge due to phase-space resonances. Furthermore, Eqs.(\ref{4})
and (\ref{5}) deliver the equations for the guiding-center
4-velocity, while (\ref{7}) implies that the physical
realizability condition is satisfied also by the guiding-center
velocity. We stress that these equations, while equivalent to
those obtained earlier \cite{Beklemishev1999}, have a simpler
structure. This follows from the adoption of the synchronous
variational\ principle here adopted which makes explicit the inner
structure and relationship existing between the gyrokinetic
variables.

\section{5. Conclusions}

In this paper we have formulated a covariant gyrokinetic approach
for single-particle dynamics, which applies to curved space-time.
The theory goes beyond the domain of validity of the treatment
previously considered by Beklemishev and Tessarotto
\cite{Beklemishev1999} and includes the case in which the electric
field can locally vanish or result much smaller than the magnetic
field. As an illustration of the theory the perturbative theory
has been developed to leading order in $\varepsilon$. For this
purpose we have adopted a synchronous variational principle based
on constrained dynamics. The resulting Euler-Lagrange equations
satisfy the required constraint demanded by the physical
realizability condition for the 4-velocity and displays the inner
relationships between the gyrokinetic variables. We have found
that to leading-order no effect of curvature of space-time appears
on the gyrofrequency at order $\varepsilon ^{1}$. This information
is relevant for plasma diagnostics in astrophysical
plasmas.\bigskip

\section*{Acknowledgments}
Research developed in the framework of MIUR (Ministero
Universit\'a e Ricerca Scientifica, Italy) PRIN Project \textit{\
Fundamentals of kinetic theory and applications to fluid dynamics,
magnetofluiddynamics and quantum mechanics}, partially supported
(P.N.) by Area Science Park (Area di Ricerca Scientifica e
Tecnologica, Trieste, Italy) and CMFD Consortium (Consorzio di
Magnetofluidodinamica, Trieste, Italy).



\end{document}